\documentclass[aps,prd,twocolumn,showpacs,superscriptaddress,groupedaddress,english,amsmath,amssymb,nofootinbib]{revtex4}  
\usepackage{graphicx}  
\usepackage{dcolumn}   
\usepackage{bm}        
\usepackage{amssymb}   
\usepackage{epsfig}
\usepackage{hyperref}
\usepackage{times}
\usepackage[usenames]{color}
\usepackage{url}
\usepackage[normalem]{ulem}
\usepackage[T1]{fontenc}
\usepackage[dvipsnames]{xcolor}

\begin{document}

\title{Distinguishing general relativity and $f(R)$ gravity with the gravitational lensing Minkowski functionals}
\graphicspath{{./gfigs/}}

\author{Chenxiaoji Ling}
\author{Qiao Wang}
\author{Ran Li}
\affiliation{National Astronomical Observatories, Chinese Academy of Sciences, 20A Datun Road, Chaoyang, Beijing 100012, China}
\author{Baojiu Li}
\affiliation{Institute for computational Cosmology, Department of Physics, Durham University, South Road, Durham DH1 3LE, UK}
\author{Jie Wang}
\affiliation{National Astronomical Observatories, Chinese Academy of Sciences, 20A Datun Road, Chaoyang, Beijing 100012, China}
\author{Liang Gao}
\affiliation{National Astronomical Observatories, Chinese Academy of Sciences, 20A Datun Road, Chaoyang, Beijing 100012, China}
\affiliation{Institute for computational Cosmology, Department of Physics, Durham University, South Road, Durham DH1 3LE, UK}

\date{\today}

\begin{abstract}

We explore the Minkowski functionals of weak lensing
convergence map to distinguish between $f(R)$ gravity and the general relativity (GR).
The mock weak lensing convergence maps are
constructed with a set of high-resolution simulations assuming
different gravity models. It is shown that the lensing MFs of $f(R)$ gravity
can be considerably different from that of GR because of the
environmentally dependent enhancement of structure formation.
We also investigate the effect of lensing noise on our
results, and find that it is likely to distinguish F5, F6 and GR
gravity models with a galaxy survey of $\sim3000$ degree$^2$ and with
a background source number density of $n_g=30~{\rm arcmin}^{-2}$, comparable to
an upcoming survey dark energy survey (DES). We also find that the $f(R)$
signal can be partially degenerate with the effect of changing cosmology,
but combined use of other observations, such as the cosmic microwave background (CMB) data, can help break this degeneracy.
\end{abstract}
\pacs{04.50.Kd,95.30.Sf,95.36.+x,98.65.Dx}
\maketitle


\section{Introduction}

It is fundamentally important to explain the
  observed accelerating expansion of the Universe
  \citep{Riess,Garnavich}. In the current understanding, this accelerating
  expansion either is driven by an exotic dark energy in the framework of
  general relativity (hereafter GR) or indicates that GR needs
  to be modified on large scales \citep{clifton}. A well-studied example
  of the latter scenario is the so-called $f(R)$
  gravity \citep{carroll}, in which the Ricci scalar $R$ in the standard
  Einstein-Hilbert action is replaced by a function $f(R)$. In most $f(R)$ models
  studied so far, the difference between $f(R)$ and $R$ remains roughly a constant
  throughout the cosmic history, therefore accelerating the expansion of the
  Universe like in the standard $\Lambda$CDM paradigm.

Although the background expansion history
  in $f(R)$ models could be practically indistinguishable from that of
  $\Lambda$CDM, the structure formation can be very different for
  these two scenarios. In $f(R)$ gravity, $df/dR$ is nontrivial and behaves
  like a dynamical scalar field, which propagates a "fifth force" between
  matter particles. The strength of this fifth force can be maximally 1/3 of that of
  Newtonian gravity, but it is usually weaker because of the well-known
  chameleon mechanism \citep{Khoury}, which strongly suppresses it in
  regions with high matter density (or deep gravitational potential). The idea
  is that any deviation from standard GR gravitational law would be "screened"
  and therefore undetectable in the solar system, in which the validity of GR has been
  confirmed experimentally to very high precision. However, it is worth stressing
  that the behavior of the fifth force in solar-like systems relies heavily on
  what is going on at much larger scales such as the Milky Way galaxy,
  its dark matter halo and beyond. Although recent works have demonstrated
  the encouraging potential of constraining $f(R)$ gravity using such systems
  \citep{sakstein1,sakstein2}, better understandings of the large-scale behavior
  of the scalar field will be needed before quantitative conclusions
  are finally drawn. In this sense, it is crucial to study the cosmological behavior
  of $f(R)$ gravity, as a means to constrain gravity using the constantly improving
  cosmological data (see \citep{lombriser2014} for a recent review).

  Previous works on this subject using statistics of
  large scale structure often compare matter power spectrum
  and correlation functions of matter distribution in GR and $f(R)$
  gravity \citep[see, e.g.,][]{oyaizu,zhao,Lia,hellwing}. In this work, we investigate the
  topological difference in the lensing convergence $\kappa$ map
  between GR and $f(R)$  universes. The lensing $\kappa$ map reflects
  the projected mass  distribution of the Universe; its topological
  information can be described using Minkowski
  functionals (MFs) \citep[][]{mecke,SchmalzingB}. In recent works, MFs
  have been extensively used to study the geometry properties of
  cosmic field (\citep{Schmalzinga, schmalzing1999, Hikage, codis,
    kerscher1998,petri}). It has been shown that lensing MFs contain significant
  information beyond other statistical quantities, e.g. the power
  spectrum \cite{kratochvil}, thus might provide a promising way to
  distinguish $f(R)$ and GR  model.

  In this paper, we construct mock lensing maps with a set
  of $f(R)$ and GR  cosmological simulations using the
  {\normalsize{E}\footnotesize{COSMOG}} \citep{Ecosmog} code, and investigate whether or not the MFs of
  lensing map can be used to distinguish different gravity models. In addition,
  we investigate the degeneracy effect between cosmic parameters and cosmic
  models.

This paper is organized as follows. In Section \uppercase\expandafter{\romannumeral 2} ,
  we briefly introduce the general $f(R)$ models and the $N$-body simulations
  used in this work. In Section \uppercase\expandafter{\romannumeral 3} 
  , we present our algorithm to calculate the MFs.
  In Section 4\uppercase\expandafter{\romannumeral 4}  we present our results. In Section \uppercase\expandafter{\romannumeral 5}  we discuss the effect of
  cosmic parameters,  and we give a summary in Section \uppercase\expandafter{\romannumeral 6} .


\section{The $f(R)$ cosmology}

\subsection{The $f(R)$ gravity model}

The $f(R)$ gravity model is a simple generalization  of
  standard $\Lambda$CDM paradigm by replacing the
  Ricci scalar $R$ in the Einstein-Hilbert action with an algebraic function of $R$. The
  modified action can be written as:
\begin{equation}
S=\int d^4x\sqrt{-g} \left\{\frac{M^2_{PI}}{2}[R+f(R)]+\mathcal{L}_m\right\},
\end{equation}

in which $M_{PI}$ is the reduced Planck mass, $M_{PI}^{-2}=8\pi G$,
$G$is Newton's constant, $g$ is the determinant of the metric $g_{\mu\nu}$
and $\mathcal{L}_m$ is the Lagrangian density for matter fields.

There is plenty of literature about the derivation and properties of the
  modified Einstein equations in $f(R)$ gravity, and here we shall not repeat
  the details. Instead, we simply present the equations that are directly
  relevant to the cosmic structure formation. These are the modified Poisson
  equation:

\begin{equation} \label{equ2}
\nabla^2\Phi = \frac{16\pi G}{3}a^2\delta\rho_m + \frac{a^2}{6}\delta R(f_R),
\end{equation}
and the equation of motion (EoM) of the scalar field $f_R\equiv df(R)/dR$:
\begin{equation}\label{equ2b}
\nabla^2f_R = -\frac{a^2}{3}\left[\delta R(f_R) + 8\pi G \delta\rho_m\right],
\end{equation}
in which
\begin{equation}
\delta R \equiv R- \bar{R},~~~\delta \rho_m \equiv \rho_m- \bar{\rho}_m.
\end{equation}
$\Phi$ denotes the gravitational potential, $\rho_m$ is the total density
of matter (cold dark matter and baryons), and an overbar denotes the background average.
$a$ is the cosmic scale factor and $a=1$ at present.

The $f(R)$ model has a GR limit, which is given by $f_R\rightarrow0$. In this limit,
the scalar field $f_R$ becomes nondynamical (identically zero); Eq.~(\ref{equ2b})
gives the GR relation $\delta R=-8\pi G\delta\rho_m$ and Eq.~(\ref{equ2}) reduces
to the standard Poisson equation:

\begin{equation} \label{equ5}
\nabla^2\Phi = 4\pi Ga^2\delta \rho_m.
\end{equation}
For general $f(R)$ gravity, on the other hand, the scalar field $f_R$
has a complicated behavior, and leads to an environmentally dependant
effective Newton's constant $G_{\rm eff}$.

As described in the introduction, local tests of gravity based on solar system
  observations put a tight constraint on any deviation from the
  Newtonian gravity.  The chameleon mechanism is introduced to evade
  the constraint by varying $G_{\rm eff}$ in different
  environments. In dense regions,  $\delta f_R$ becomes negligible, and
  one has

 $\delta R(f_R) \approx -8\pi G\delta \rho$,
  thus Eq.~({\ref{equ2}}) returns to the GR equation,
  Eq.~({\ref{equ5}}). In underdense environments, the $\delta
  R(f_R)$ term in Eq.~(\ref{equ2}) becomes small and the Eq.~(\ref{equ2})
  turns into:

\begin{equation} \label{equ6}
\nabla^2\Phi = \frac{16}{3}\pi Ga^2\delta \rho,
\end{equation}
where the effective Newton's constant is enhanced by a factor of $1/3$ ($G_{\rm eff}=4G/3$)
compared to its value in dense environments.

Note that the maximum enhancement of $G$ in $f(R)$ gravity is always $1/3$,
independent of the functional form of $f(R)$. $f(R)$, on the other hand, determines
how $G_{\rm eff}$ changes from G to $4G/3$ when environmental density changes.
Therefore, the form of $f(R)$ is crucial for a given model. To date, various $f(R)$
functions have been designed to explain the accelerated cosmic expansion while
evading solar system constraints, of which the most wellstudied is the one proposed by \citep{hu}:
\begin{equation}
f(R)=-M^2\frac{c_1(-R/gM^2)^n}{c_2(-R/M^2)^n+1},
\end{equation}
where $M^2 \equiv H_0^2\Omega_m$ with $H_0$ the Hubble constant,
$\Omega_m$ is the matter density parameter, and $n$ is an integer parameter
which is normally set to 1, though other values have been studied as well.
To match\footnote{We note that this is an approximate match, with a error
of order $f_{R}$, which is practically too small to be observable.
As mentioned earlier, it is possible to have an exact $\Lambda$CDM
expansion history in $f(R)$ gravity, but the corresponding form of $f(R)$
is more complicated.} the expansion of a standard $\Lambda$CDM universe,
the dimensionless parameters $c_1$ and $c_2$ should satisfy:
\begin{equation}
\frac{c_1}{c_2}=6\frac{\Omega_{\Lambda}}{\Omega_m},
\end{equation}
where $\Omega_{\Lambda}$ is the current dark energy density
parameter.

In any reasonable cosmological model,
we have $-\bar{R}\gg M^2$, and so $\bar{f}_R$ can be
simplified as:
\begin{equation}
f_R\simeq -\frac{nc_1}{c_2^2}\left(\frac{M^2}{-R}\right)^{n+1}.
\end{equation}
Therefore, the model can be described by two free parameters, $n$ and
$c_1/c_2^2$, and the latter is determined by $f_{R0}$, the value of
$f_R$ today.


\subsection{Numerical simulations}

Numerical simulations used in this study include
  three high resolution cosmological  $N$-body simulations, two of
  which assume $f(R)$ gravity and one assumes $\Lambda$CDM.
  For two different $f(R)$ simulations, we have fixed the model
  parameter $n$ to be 1 but varied the parameter $f_{R0}$ by $|f_{R0}| =
  1.289 \times 10^{-5}$ and $1.289\times10^{-6}$, which will hereafter
  be referred to as F5 and F6 model, respectively. All simulations
  evolve $1024^3$ particles in a 250 $h^{-1}$Mpc cubic volume and start
  from exactly the same initial conditions at $z=49.0$. The simulations were performed
  with the adaptive mesh refinement code {\sc Ecosmog} \cite{Ecosmog}. The
  cosmological parameters assumed to generate initial conditions are
  $\Omega_m=0.267$,  $\Omega_\Lambda=0.733, h=0.71,n_s=0.958$ and
  $\sigma_8 = 0.801$, in which $h=H_0/(100{\rm km/s/Mpc})$, $n_s$ is the spectral
  index of the primordial power spectrum and $\sigma_8$ is the rms density
  fluctuation within spherical tophat windows of radius 8 $h^{-1}$Mpc.
  In this work, we place source galaxy at $ z = 1$ and use
  the snapshot at $z \approx 0.1$ to construct lensing $\kappa$ maps.

\section{The Minkowski functionals (MFs) of  Weak Lensing $\kappa$ map}

\subsection{Weak lensing convergence map}

Weak lensing observations measure small distortions on the shapes of background
galaxies, which can be used to generate convergence $\kappa$ map. The convergence
map $\kappa(\mathbf{x})$ is related to projected density map $\Sigma(\mathbf{x})$ as:
\begin{equation}\label{eq:eps}
	 \kappa(\mathbf{x})=\frac{\Sigma(\mathbf{x})}{\Sigma_{cr}},
\end{equation}
with the critical surface density
\begin{equation}\label{eq;eps}
	\Sigma_{cr}=\frac{c^2}{4 \pi G}\frac{D_s}{D_lD_{ls}},
\end{equation}
in which $D_{ls}$ is the angular diameter distance between source
galaxies and the lens, and $D_l$ and $D_s$ are the angular diameter
distances from the observer to the lens and to the sources. $c$ is the speed of light.

To generate a theoretical convergence map, we project particles in
the whole simulation box onto a plane. Next we employ the cloud in
cell (CIC) method to project dark matter particles to a $5000^{2}$
grid surface density map. On average, there are about $43$ particles
on each grid, the grid separation is about $50 h^{-1} kpc$. Then we convert the
surface density map to convergence map by assuming our lens plane to be at
$z=0.1$, and all source galaxies at redshift $z=1$. The
total sky area of our mock lensing observation is about 3000
degree$^2$, comparable to forthcoming dark energy surveys (e.g., LSST
\citep{lsst} and Euclid \citep{euclid}).

In real observations, the intrinsic ellipticity of source galaxies
introduces noise to the convergence map. The Gaussian smoothing is
often adopted to suppress the small scale noise. The uncertainties of
a smoothed $\kappa$ map are specified by the number density of source
galaxies, $n_g$, and the smoothing aperture size $\theta_G$. van
Waerbeke 2000 \citep{van2000} shows that the noise can be approximated
by a Gaussian distribution with rms:
\begin{equation}
\label{eq:noise}
	\sigma^2_{noise}=\frac{\sigma^2_\epsilon}{4 \pi \theta^2_Gn_g},
\end{equation}
where $\sigma^2_{\epsilon}$ is the rms amplitude of the source
intrinsic ellipticity distribution.

To simulate a more realistic convergence map, we first smooth our
convergence map with the Gaussian window. We then add the noise
resulting from intrinsic ellipticity of source galaxies using
Eq.(\ref{eq:noise}). Following \citep{Hamana}, we set
$\sigma_\epsilon=0.4$.

To investigate the effect of smoothing scale on our results, we adopt
three different smoothing scale $\theta_G=0.5$, $1$ and $5$ arcmin. For $n_g$,
we adopt two values: $n_{g1} = 30~{\rm arcmin}^{-2}$ for upcoming surveys
such as DES and $n_{g2} = 100~{\rm arcmin}^{-2}$ for future more ambitious
surveys.

\subsection{Minkowski functionals}

Minkowski functionals provide morphological
  statistics for any given smoothed random field  characterized by a
  certain threshold $\nu$. Compared with traditional power spectrum
  methods, MFs contain not only information of spatial correlation of a
  random field, but also information of topology and object
  shapes. For a $\mathbb{R}^n$  field one can get $n+1$ MFs $V_i$.
  Weak lensing convergence map is a two-dimensional field, thus 3 MFs
  can be defined, namely $V_0$,  $V_1$, and $V_2$.

For a smoothed field $u(\mathbf{x})$ in a 2D space,
we define the area $Q_\nu$ and boundary
$\partial Q_\nu$  to be:
\begin{displaymath}
Q_\nu \equiv \lbrace \mathbf{x} \in \mathbb{R}^2 | u(\mathbf{x}) > \nu \rbrace, \ \ \
\partial Q_\nu \equiv \lbrace \mathbf{x} \in \mathbb{R}^2 | u(\mathbf{x}) = \nu \rbrace.
\end{displaymath}

Then, MFs can be written as follows:

\begin{equation}
\label{MF_V0}
V_0(\nu) =\int_{Q_\nu} d\Omega,
\end{equation}
\begin{equation}
\label{MF_V1}
V_1(\nu) = \int_{\partial Q_\nu} \frac{1}{4} dl,
\end{equation}
\begin{equation}
\label{MF_V2}
V_2(\nu) = \int_{\partial Q_\nu} \frac{1}{2 \pi}\kappa_cdl.
\end{equation}
$V_0$ is the area of $Q_\nu$, $V_1$ is the total
boundary length of $Q_\nu$ and $V_2$ is the integrated geodesic
curvature $\kappa_c$ along the boundary.

We follow the method described in \cite{Lim,kratochvil} to calculate the MFs from the
pixelated maps. On each grid, we calculate:
\begin{equation}
\mathcal{I}_0 (\nu,p_j) = \Theta(u-\nu),
\end{equation}
\begin{equation}
\mathcal{I}_1 (\nu,p_j) = \frac{1}{4}\delta(u-\nu)\sqrt{u_{,x}^2+u_{,y}^2},
\end{equation}
\begin{equation}
\mathcal{I}_2 (\nu,p_j) = \frac{1}{2 \pi}\delta(u-\nu)\frac{2u_{,x}u_{,y}u_{,xy}-u^2_{,x}u_{,yy}-u^2_{,y}u_{,xx}}{{u^2_{,x}+u^2_{,y}}},
\end{equation}
where $u_x, u_y$ are the two partial derivatives of $u(\bf x)$.
The numerical MFs of $V_i$ can be computed by summing integrands over all pixels:
\begin{equation}
\label{NMFs}
V_i(\nu)=\frac{1}{N_{pix}}\sum_{j=1}^{N_{pix}} \mathcal{I}_i(\nu,p_j),
\end{equation}
In the above, $\Theta$ is the Heaviside step function. For the bin width $\Delta_\nu$,
the delta function can be numerically calculated
as follows:
\begin{equation}
\delta_N(\nu)=(\Delta\nu)^{-1}[\Theta(\nu+\Delta\nu/2)-\Theta(\nu-\Delta\nu/2)].
\end{equation}

Note that the numerical MFs, Eq.(~\ref{NMFs}), is actually the surface
density of Eqs.(~\ref{MF_V0}), (\ref{MF_V1}) and (\ref{MF_V2}). In what
follows, we refer to both of them as MFs and notation $V_i$.

\section{Results}

\begin{figure}[!ht]
\centering
\includegraphics[width=3.25in]{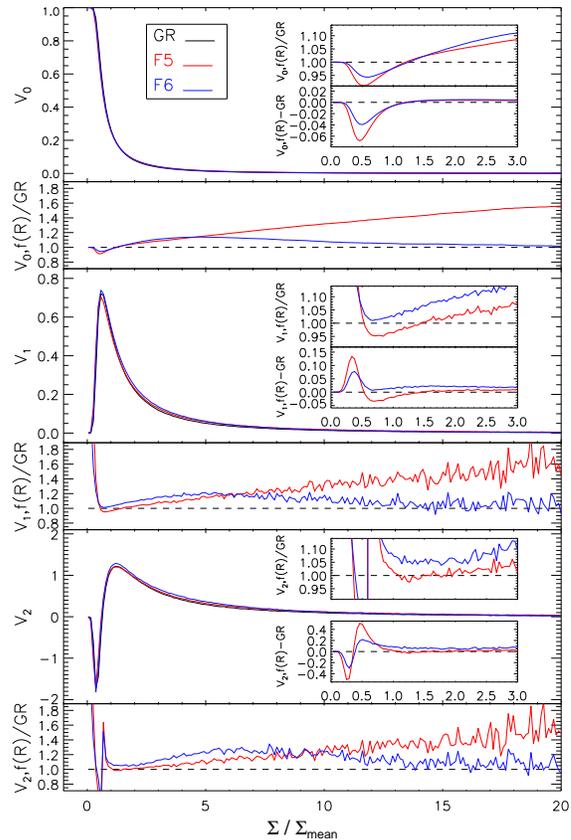}
\caption{\label{fig:theory}
  The MFs of surface density maps as a function of surface density
  normalized to mean of the universe $\Sigma_{\rm mean}$ for our $f(R)$
  and GR simulations. The black solid, red dashed, and blue
  dotted lines represent result for GR, F5 and F6 simulation,
  respectively. Interior small figures each panel show the ratios
  between GR and $f(R)$ simulations and the residuals as a function of
  surface density.}
\end{figure}

In Fig.~\ref{fig:theory}, we show the MFs of
  surface density from our simulations. The MFs are plotted as
  functions of surface density in unit of mean surface density, $\Sigma_{\rm mean}$.
  The overall shapes of MF curves of the $f(R)$ and GR models are
  similar. However, the amplitude of MFs of the $f(R)$ surface
  density map is higher at $\Sigma/\Sigma{\rm mean}>2$. For the F6
  case, $V_0$ is  $\sim 10\%$ higher than that of GR model at
  $\Sigma/\Sigma_{\rm mean} \sim 3-5$, while in denser
  regions ($\Sigma/\Sigma_{\rm mean}>15$) the $V_0$ of both models
  are almost identical. On the other hand, the difference in
  $V_0$ between F5 and GR increases with $\Sigma/\Sigma_{\rm mean}$
  and persists to larger density.  At $\Sigma/\Sigma_{\rm mean} \sim
  20$,  the $V_0$ of F5 model is about 60\% larger than that
  of GR model. The $V_1$ and $V_2$ of F5 and F6 models show
  similar trends.

The apparent differences shown here reflect the
  environmentally dependent structure formation in universes with
  different gravity theories. As is shown in Reference ~\citep{Lib},
  compared with the GR universe, there are more massive halos and larger
  size voids in $f(R)$ models because of the enhanced gravity in low
  density environments. As a result, the surrounding regions (including
  the filaments) of dark matter halos are denser in $f(R)$
  gravity than in GR. Therefore, in $f(R)$ models,  $V_0$, which represents the
  area of regions with density higher than certain threshold, is
  smaller than that of GR in the low density
  regions ($\Sigma/\Sigma_{\rm mean}<1$), but is larger at relatively
  high density regions. For F6, in very high density regions, the
  chameleon screening ensures that both the gravity and MFs are
  similar to the results in GR.

The difference in $V_1$ and $V_2$ can also be
  explained in the similar way. However, unlike $V_0$, $V_1$ and
  $V_2$ encode additional information on topology (which describes
  connectivity) of the  $\kappa$ map. As an example, the turn over trend
  in the lower panel of  $V_1$ indicates transition of the topology of
  $\kappa$ map from the isolated  halo dominated case to the voids
  dominated one.

In real observations, noise resulting from the
  intrinsic ellipticity distribution of galaxies contaminates the
  lensing $\kappa$ map. Gaussian smoothing is usually adopted to
  suppress the noise; however, it will also mix the  MFs of different
  density thresholds.

\begin{figure*}[!ht]
\centering
\includegraphics[width=6.5in]{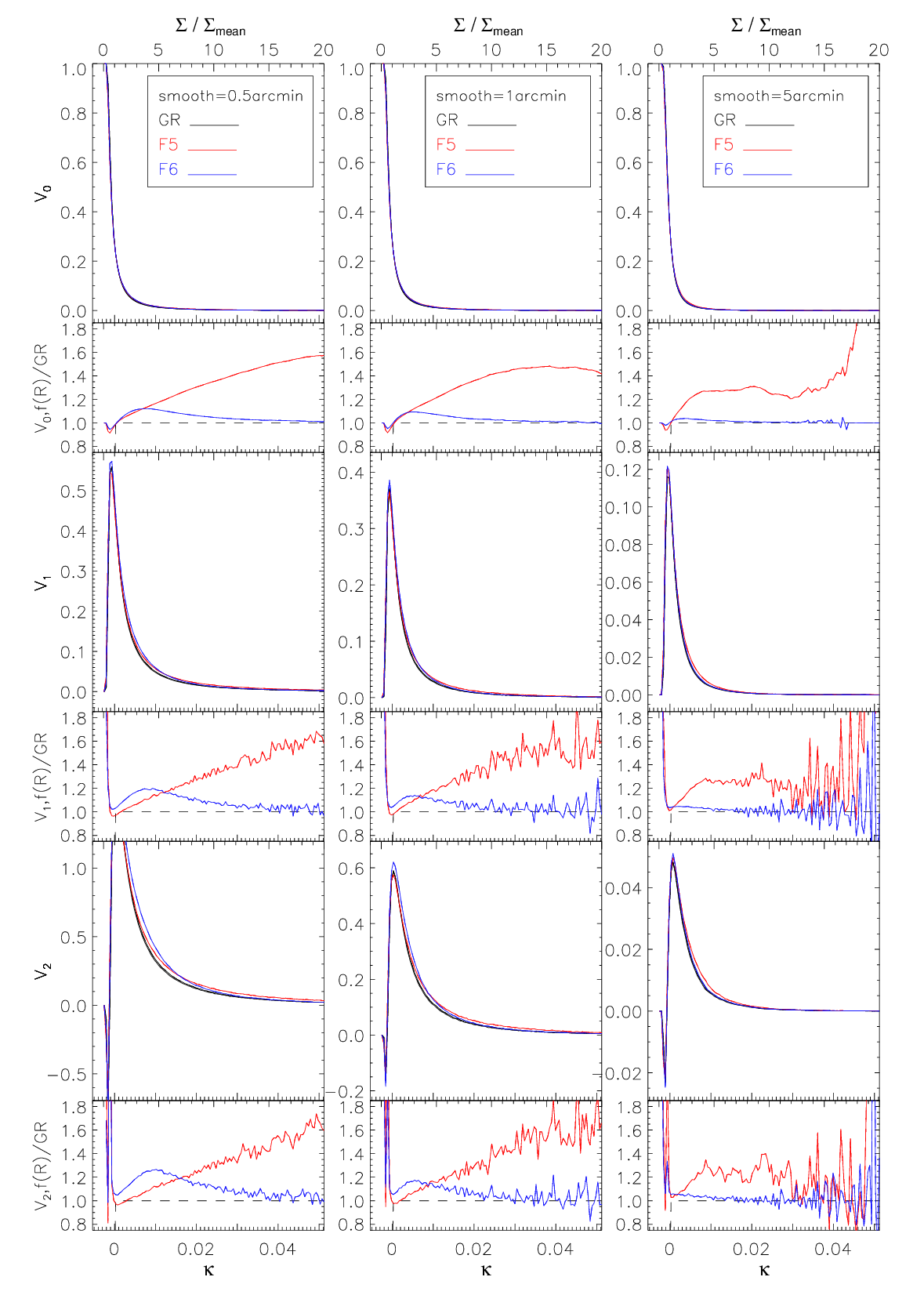}
\caption{\label{fig:theorysmooth}
  The MFs of the noiseless $\kappa$
  maps as a function of $\kappa$ (lower axis) and
  $\Sigma/\Sigma_{mean}$ (upper axis). The black solid, red dashed,
  and blue dotted lines represent result of GR, F5 and F6 models
  respectively.  The panels from left to right show results with
  different smoothing scales: $0.5'$ , $1'$ and  $5'$, respectively.}
\end{figure*}

In Fig.~\ref{fig:theorysmooth}, we show the MFs of
  the simulated $\kappa$ maps without taking into account the noise. We
  apply smoothing to the map  with different smoothing scales,
  $\theta_G=0.5$, $1$ and $5$, respectively. Ref.~\citep{Hamana}
  claimed that for cluster survey, the best
  smoothing scales is $\sim 1arcmin$  . We find that the amplitudes of MFs at high
  density regions decrease significantly while the Poisson noises
  increase dramatically at the same region.  This is because the
  smoothing procedure reduces total area of high density region. We
  conclude that a small smoothing scale is better for measuring MFs.

\begin{figure*}[!ht]
    \centering
	\includegraphics[width=6.5in]{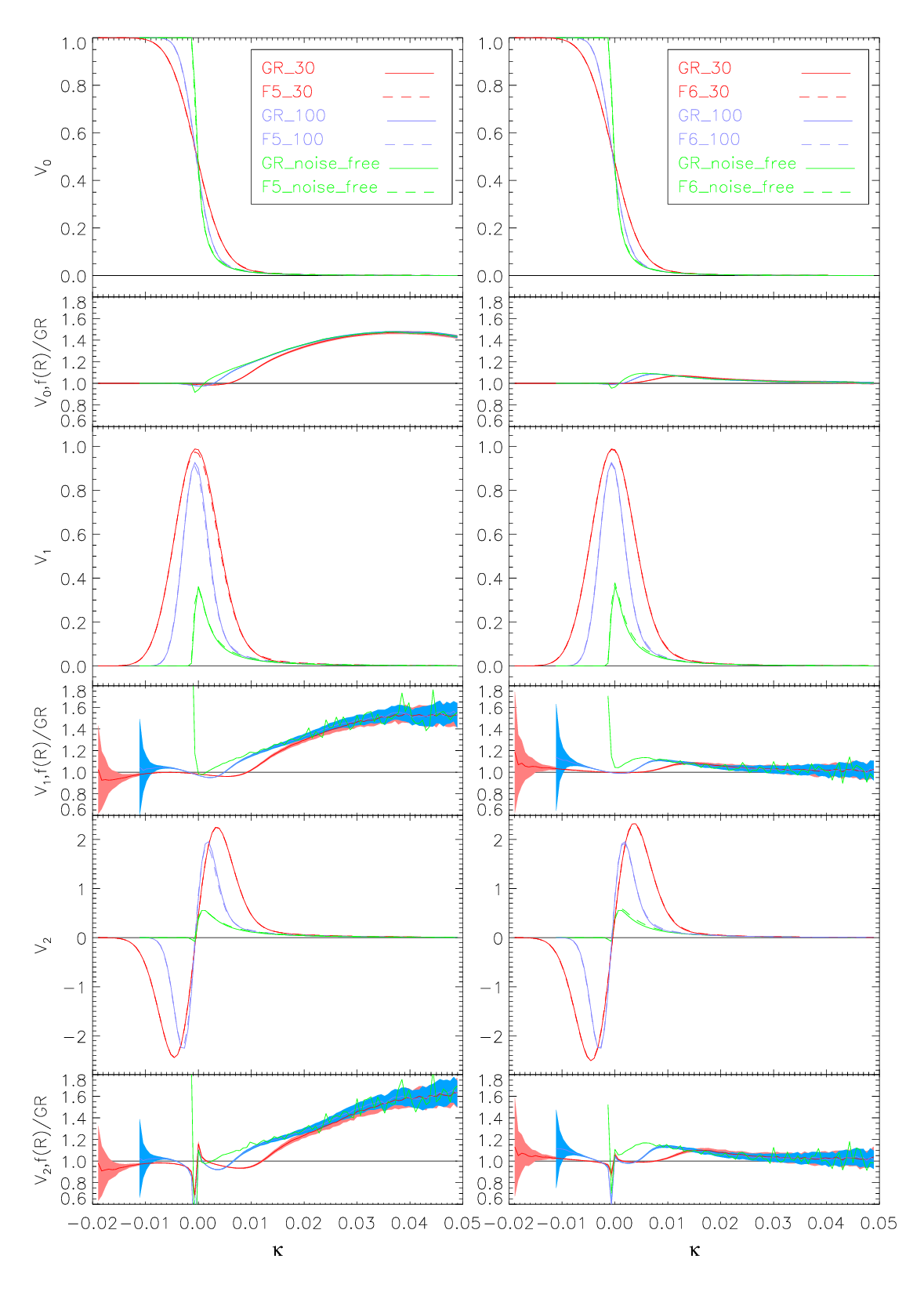}
	\caption{\label{fig:mock}
	Comparisons of the predicted MFs of lensing
	$\kappa$ maps between the $f(R)$ and GR simulations for three different
	source number densities, $n_g=30$, $100$ and infinity (noise-free).
	In all cases the smoothing scale is taken to be $1'$.
	The left panel show comparison between F5 and GR models, while the right panel
	show comparison between F6 and GR models. The shaded regions show
	standard deviation of MFs from 100 mocked lensing maps.}
\end{figure*}

We show MFs of the $\kappa$ map in
   Fig.~\ref{fig:mock}  by taking into account the noise. Here we
   adopt two different noise cases, $n_g=30$ and $n_g=100$, in order to
   investigate effects of different noise levels on our results.
   For comparison we also include the case without noise.
   In the figure, we use a smoothing scale of $1'$ and generate $100$
   maps using different background noises which sharing the same
   standard  deviation $\sigma_{noise}$. The shaded regions show the
   standard deviation of MFs, which is an estimation of the noise
   level.  We note that the noise map due to intrinsic galaxy shapes can be
   approximated with a Gaussian map, which migrate into  MFs, thus
   suppressing the difference between $f(R)$ and GR models. However, this effect is
   less important in dense regions ($\kappa>0.015$). Therefore, it is
   still possible to distinguish the F5, F6 and GR models in high
   density regime. It is also interesting to see that even for the higher noise
   level, where $n_g=30~{\rm arcmin}^{-2}$, the difference among different
   gravity models is still much larger than the observational lensing
   noise, indicating that weak lensing MFs can be a powerful tool to
   distinguish the $f(R)$ and GR models with upcoming galaxy surveys.

   We further tested the evolution of lensing MFs with redshift.
  	 For this we fixed the source galaxy at $z=1$ and employed four snapshots
  	 at different lens redshifts ($z=0.8$, $0.5$, $0.2$ and $0.1$ respectively).
  	 Figure 4 presents the results, where for simplicity we have used the same
  	 source galaxy density ($n_g=30$) and smoothing scale ($1'$). The amplitudes
  	 of $V_1$ and $V_2$ of both GR and $f(R)$ gravity increase with time,
  	 and the relative differences of MFs between F5 and GR grow from 15\%
  	 (at $z=0.8$) to $50\sim60\%$ (at $z=0.1$). The same trend is found for
  	 F6 but the deviation from GR is much weaker. The results suggest that
  	 even with a more realistic line-of-sight integration to fully account
  	 for the matter distribution, we expect the model difference to be still present.
  	 Note that such an integration would somewhat distort the results
  	 and suppress the non-Gaussianity of the signal.

\begin{figure*}[!ht]
   	\centering
   	\includegraphics[width=6.5in]{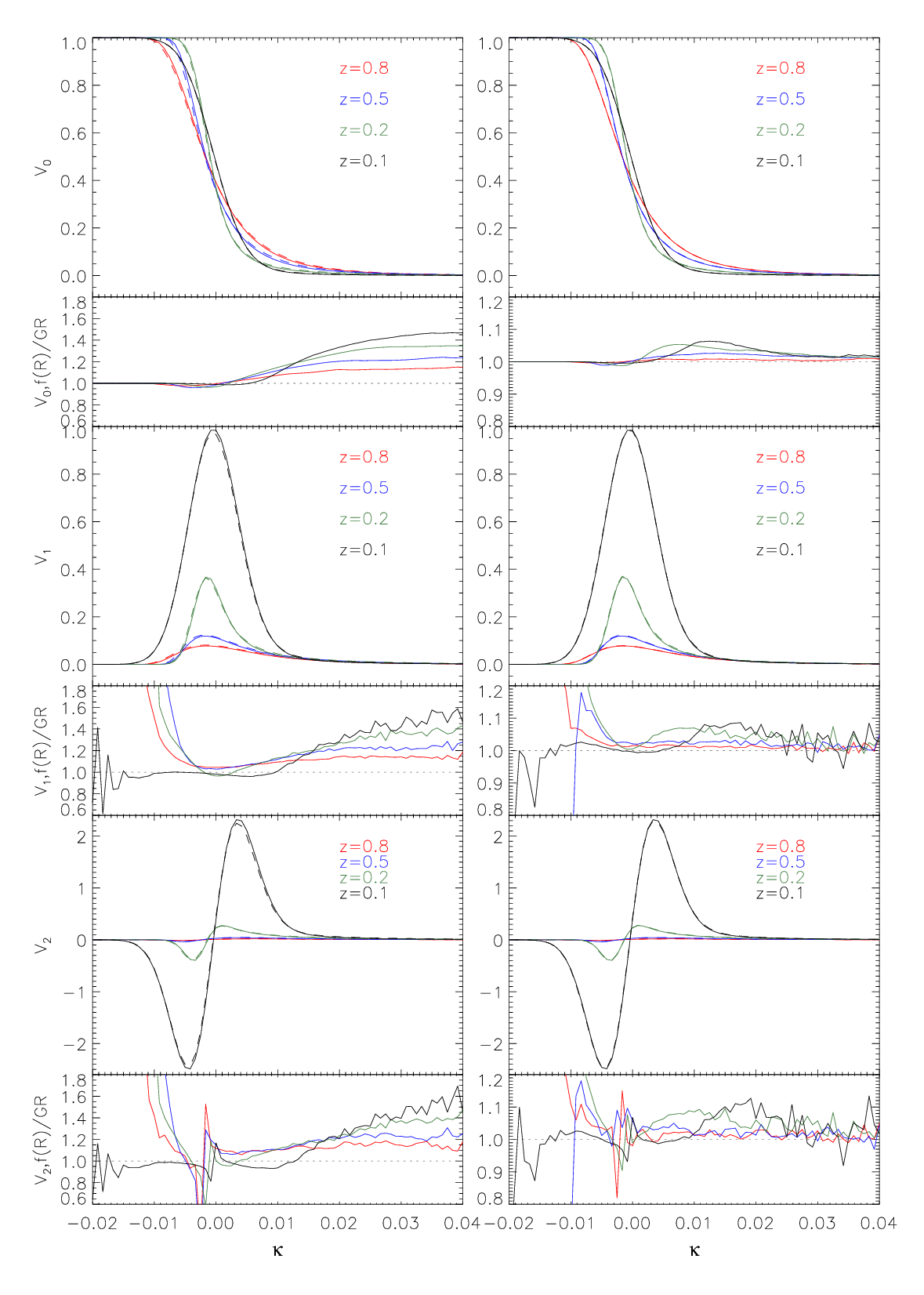}
   	\caption{\label{fig:redshift} Lensing MFs at $z=0.8$, $0.5$, $0.2$ and $0.1$,
   	 with the same smoothing scale ($1'$) and source number density ($n_g=30$).
   	 The left panels show the comparison between F5 and GR, while the right panels
   	 are the comparison between F6 and GR. The dashed (solid) lines are results
   	 for $f(R)$ gravity (GR).}
\end{figure*}

\section{The effect of cosmic parameters}

 In this work, we have mainly focused on the difference between
	the MFs for $f(R)$ gravity and GR, but note that this signal could in
	principle be degenerate with the effect of changing cosmology \cite{shirasaki2014}.
	To gain a rough idea of this degeneracy, we have employed two additional
	simulations: the Millennium simulation (MS) and a MS-W7 simulation.
	These simulations are identical on the simulation box and mass resolution, but with the
	cosmology changed from WMAP1 \cite{springle} to WMAP7 \cite{guo}. Because these two simulations  
	are	carried out by different simulation codes from $f(R)$ and GR runs, this code changing could 
	bring about another effect in the MFs' measurement. We think it is better to check the effect of 
	changing cosmology and changing gravity models individually, we present the comparison between 
	two MS runs and comparison between different gravity models separately.

In Fig.~5, we compare the ratios of MFs for F5/GR (dotted lines),
	F6/GR (dashed lines) and WMAP1/WMAP7 (solid lines). The results of the noise-free
	case are shown in the left panels, while the right panels are the results assuming
	a source galaxy number density of 30. We find that the change of cosmology from
	WMAP7 to WMAP1 can have a similar impact as having F5 instead of GR as the gravity model.
	However, with the precision of current observations, the WMAP1 and WMAP7 cosmologies
	can be distinguished by using CMB data alone. In the cases of F6 and F5,
    the CMB power spectra are practically the same as GR predictions with the same
     cosmological parameters.Therefore, the CMB constraints on those cosmological parameters
    can be used to break the degeneracy above.

\begin{figure*}[!ht]
	\centering
	\includegraphics[width=6.5in]{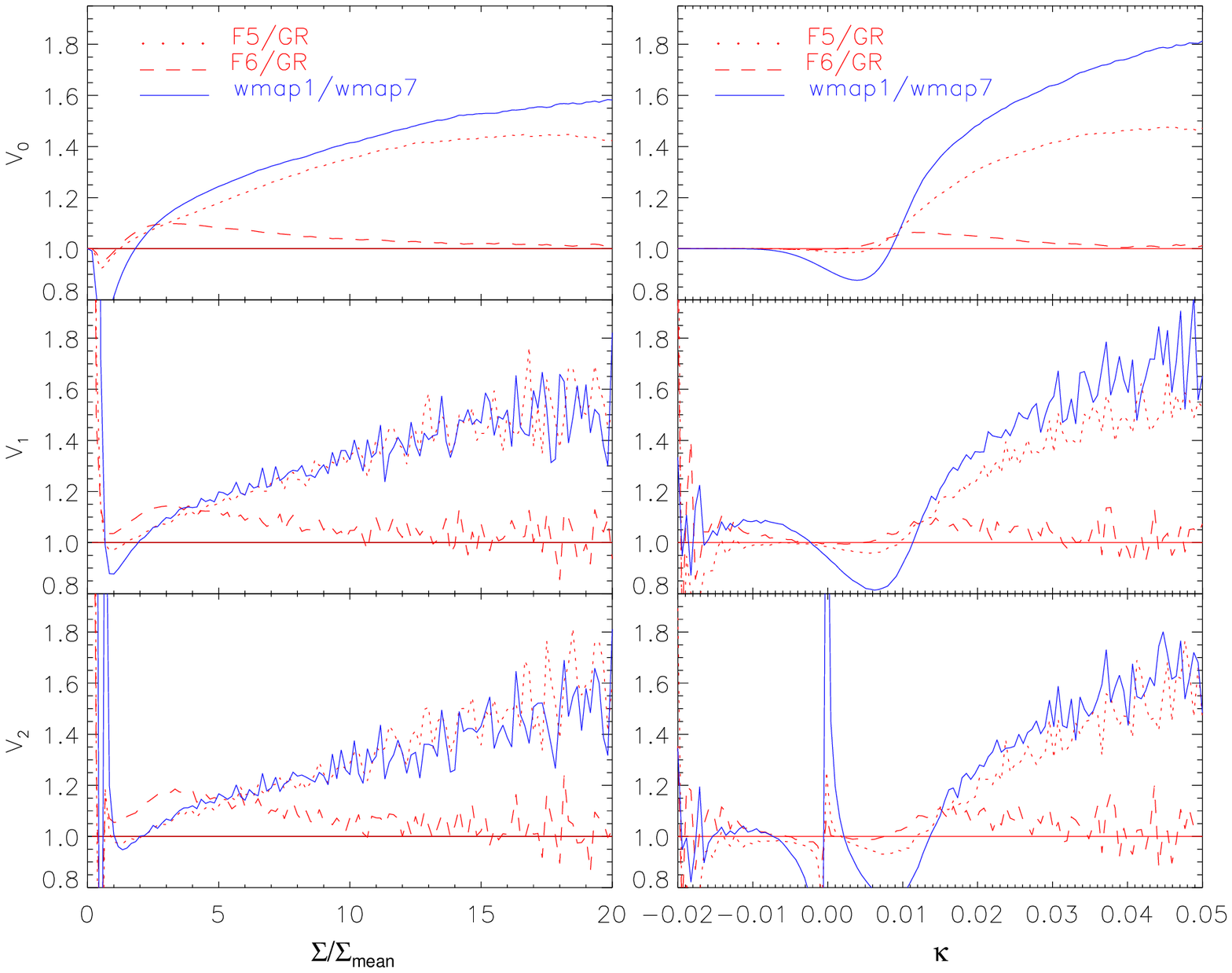}
	\caption{\label{fig:para}
	  Comparison of the effects on MFs from changing cosmological parameters
	  from WMAP7 to WMAP1 and changing the gravity model (from GR to $f(R)$ gravity).
	  The left panels show the results without noise, while the right panels assume a
	  source number density $n_g=30$. In all cases the smoothing scale is $1'$.}
\end{figure*}

\section{Summary}

In this work, we make use of high-resolution
  $f(R)$ (F5, F6) and GR simulations to generate mock lensing $\kappa$
  map by taking into account different noise levels. We find  that due
  to environmental dependent nature of $f(R)$ gravity, the MFs of
  their $\kappa$ maps show considerable deviation from the GR case. We
  also investigate the effect of lensing noise on our results, and
  find that while noise due to limited background source density
  induce pollution to the $\kappa$ map, the difference between F5, F6 and GR
  gravity models can still be distinguished with a survey of $\sim 3000$
  degree$^2$ area and with a background source number density
  $n_g=30~{\rm arcmin}^{-2}$. Such a requirement can be achieved by upcoming
  lensing surveys. We compared the effect of changing cosmological
  parameters and found that it can partly degenerate with the signal found in
  modified gravity. However, combined use of CMB data can help to break this degeneracy.
  Our results hence suggest that the  MFs of lensing $\kappa$ map
  will be a powerful tool to study the nature of gravity in the future.

\section*{Acknowledgements}

We acknowledge support from the National Natural Science Foundation of China ({\small NSFC}) Grant
  No. 11303033, No.11103011, No.11373029, No.11390372, No.11403035,
  No. 11261140641, the Strategic Priority Research Program "The
  Emergence of Cosmological Structure"  of the Chinese Academy of
  Sciences (CAS) (No. XDB09000000), {\small MPG} partner Group family.
  R.L. acknowledges the support from Youth Innovation Promotion Association of CAS.
  J.W. acknowledges supports from the Newton Alumni Fellowship,
  the 1000-young talents program, the 973 program grant No.
  2013CB837900, No.2015CB857005, the CAS grant(No. KJZD-EW-T01). L.G.
  acknowledge a Science and Technology Facilities Council ({\small STFC}) Advanced Fellowship, as well as
  the hospitality of the Institute for Computational Cosmology (ICC) at
  Durham University. We thank the anonymous referee for comments which helped us to improve the paper.
  The simulations in this work used the DiRAC
  Data Centric system at Durham University, operated by the ICC
  on behalf of the STFC DiRAC HPC Facility
  (www.dirac.ac.uk). This equipment was funded by BIS National
  E-infrastructure Capital Grant NO.ST/K00042X/1, STFC Capital Grant
  NO.ST/H008519/1, and STFC DiRAC Operations Grant NO.ST/K003267/1 and
  Durham University. DiRAC is part of the National E-Infrastructure.

\bibliographystyle{unsrt}
\bibliography{mf_fr}

\end{document}